\documentstyle[aps]{revtex}

\begin{document}
\title{Mode - mode coupling and primordial fluctuations}
\author{Esteban Calzetta and Mariana Gra\~na}
\address{IAFE and Physics Department,\\
University of Buenos Aires, Argentina}
\maketitle

\begin{abstract}
We consider the influence of mode - mode coupling in the inflaton field on
the spectrum of primordial fluctuations. To this end, we formulate a
phenomenological model where the inflaton fluctuations are treated as a
fluid undergoing turbulent motion. Under suitable assumptions, it is
possible to estimate the size and scale of fluctuations in velocity, which
upon reheating induce corresponding fluctuations in the radiation energy
density. For De Sitter inflation the resulting spectrum is scale invariant
on all scales of interest. The amplitude of the resulting spectrum is
compatible with known observational limits. This suggests that the
hypothesis of a extremely weakly coupled inflation could be relaxed without
affecting the predictions of the model. Principal PACS No 98.80.Cq;
additional PACS nos: 04.62.+v, 11.10.Wx, 47.27
\end{abstract}

\section{Introduction}

The object of this paper is to present models of fully nonlinear
fluctuations during the inflationary era conducing to a scale invariant
primordial density contrast spectrum with an amplitude consistent with COBE
observations.

Inflationary models were originally introduced as a solution for the
so-called puzzles of standard hot Big Bang cosmology \cite{Peebles}, namely
the horizon, flatness and photon to baryon ratio problem \cite{guth} \cite
{abbottpi}. A period of exponential expansion of the Universe was
postulated, whereby the ratio of the Universe increased by a factor $e^{60}$
or so, followed by a period of reheating, during which the temperature of
radiation was raised up to a final value of $10^9GeV$ or more (we assume
units with $\hbar =c=k=1$) \cite{BdV}. From then on, cosmic evolution
followed the lines of standard cosmology \cite{weinberg3'}.

Although many implementations of inflation have been proposed, most
attention has been devoted to the simplest scenario, which was Linde's
chaotic inflation with a single inflaton field \cite{linde}. In this model,
inflation is powered by a scalar field $\phi ,$ the inflaton, slowly rolling
down an effective potential $V\left( \phi \right) .$ We shall simplify the
model further by assuming the Universe is described by a spatially flat,
Friedmann - Robertson - Walker geometry. Inflation begins when the effective
potential becomes the dominant form of energy density, and ends with the
decay of the inflaton onto radiation. During inflation the potential energy $%
V\left( \phi \right) $ acts as an effective cosmological constant. It can be
said that no satisfactory contending explanation of the cosmic puzzles is
available \cite{TWH}.

Soon after the original proposal, it was realized that inflation could
perform a subtler task: to provide a framework for explaining the origin of
primordial density fluctuations \cite{nuffield}. Quantum fluctuations of the
inflaton field distort the reheating surface, inducing a primordial density
contrast (see \cite{Peebles}, we shall review this argument in greater
detail below)

\begin{equation}
\frac{\delta \rho }\rho \sim \frac H{\dot \phi }\delta \phi
\label{pridencon}
\end{equation}
$H$ being the inflationary Hubble parameter. A similar argument shows that
gravitational waves are also being created, with an amplitude $h\sim H/m_p$.
The validity of Eq.(\ref{pridencon}) has been corroborated by several
different methods \cite{BMPR} (for a dissenting viewpoint, see \cite
{Grischuck}).

In order to obtain a concrete prediction from Eq.(\ref{pridencon}) we must
estimate the quantum fluctuations $\delta \phi $. The usual approach treats
these fluctuations as a free field (for example, in the seminal paper by
Starobinsky \cite{Starobinsky}) in its (De Sitter invariant) vacuum state or
very close to it \cite{DSvacua}. Then a simple calculation within quantum
field theory in curved spaces allows us to evaluate the $\delta \phi $ in
Eq.(\ref{pridencon}) as $\delta \phi \sim H$ \cite{BirDv}. All quantities in
the right hand side of Eq.(\ref{pridencon}) are evaluated as the relevant
mode leaves the horizon. There are well defined ways to relate the spectrum
of scalar and tensor primordial fluctuations to a given potential, which are
generally described as the potential reconstruction program \cite
{reconstruction}.

As our understanding of the formalism and the complexity of observational
data progresses \cite{COBE}, it becomes clear that simple potentials such as
single powers of $\phi $ are not rich enough to account for observations.
Thus the potential reconstruction program generally assumes more complex
functional forms, depending upon several parameters \cite{reconstruction}.
In these more general potentials their higher derivatives may not be
negligible, leading to nonlinear interactions between fluctuations. In such
a case, the usual way of estimating the primordial density contrast would be
invalid. The relevance of non linear fluctuations was already taken up in
Ref. \cite{dursak}. Also several authors \cite{Morikawa,Matacz,CH95,CG97}
have considered models where fluctuation self interactions were treated
perturbatively.

The main goal of this paper is to develop ways of estimating the primordial
density contrast in chaotic inflation models without presupposing that
couplings among fluctuations are negligible. Of course, working out the
structure of quantum fluctuations on even simple states of fully nonlinear
quantum field theory is a daunting task\cite{weinberg96}. However, a simpler
alternative is available, namely, the application of hydrodynamics to
describe the macroscopic behavior of quantum fluctuations. This is possible
because these fluctuations, as far as it is relevant to our present
concerns, may be described by a c-number energy momentum tensor subject both
to the usual conservation laws and the Second Law of thermodynamics \cite
{BirDv}. There is therefore an equivalent fluid description, consisting of a
classical fluid whose energy momentum tensor and equation of state reproduce
the observed ones for the quantum fluctuations. Solving the dynamics of this
equivalent fluid yields answers to all relevant questions concerning the
behavior of the actual quantum fluctuations. In this paper we shall not
press the issue of whether the equivalent fluid is anything more than a
convenient computational device.

Along with the assumption of free inflaton fluctuations, we must question
whether these fluctuations are in their vacuum state. This assumption
usually rests on the so called {\it cosmic no hair theorem, }which states
that any reasonable initial quantum state for the field will quickly relax
to a De Sitter invariant state, or its closest available analogue \cite{CNHT}%
. The application of this theorem in models with a hundred or more
e-foldings of inflation is justified, but these models must be rejected on
the grounds that they also predict a value of the density parameter $\Omega $
unacceptably close to one (see Appendix and Ref. \cite{Grishchukdur}; it is
also possible to develop inflationary models in open universes, see Ref. 
\cite{Cohn}). When the duration of inflation is close to minimal, it becomes
likely that some large but observable scales will leave the horizon before
the cosmic no hair theorem is able to operate. It is important therefore to
set a realistic initial condition such as a Planckian distribution of
particles with temperature $T$. The relevance of thermal effects in
inflation was already pointed out (in a different context) in ref. \cite
{Berera}.

An immediate consequence of energy momentum conservation and the Second Law
is that when velocities are low, the phenomenological fluid may be described
within the Eckart theory of dissipative fluids \cite{Weinberg,Hu} (for an
analysis of the limitations of Eckart's theory see \cite{Geroch}). It
follows that it obeys a continuity equation and a curved space time Navier -
Stokes equation. The model is then defined by giving the equation of state
and the viscosity of the equivalent fluid. The advantage we gain is that
these are features that can be computed locally. As far as the relevant
scales are much below the curvature radius, it is possible to use for them
their standard flat space time values. At high temperatures we obtain the
equation of state for radiation, $p=(1/3)\rho $, and a dynamic viscosity $%
\eta \sim T^3$. Since the speed of sound, being close to light speed, is
much higher than the characteristic speed of the fluid, the flow may be
considered incompressible. There will be fluctuations in velocity,
nevertheless, and these are the ones responsible for density fluctuations,
as it is well known \cite{Peebles}.

As we shall show below, conditions in the early stages of inflation are such
that, for generic initial conditions, the flow of the equivalent fluid is
highly turbulent, meaning that the corresponding Reynolds number is well
over a thousand. The possible role of turbulence in the formation of cosmic
structures has been studied in some detail even before inflationary models
were introduced (see \cite{PeeblesLSS,Raychaudhuri} for a textbook account).
These early attempts were abandoned because there were no natural mechanisms
for the production of primordial turbulence, and the density contrast
predicted was generally too high. After inflation was proposed, the matter
was taken up again by Goldman and Canuto \cite{GC}, who studied longitudinal
turbulence excited by density fluctuations in the radiation and matter
dominated eras. Our work should not be understood as a continuation of this
line of research, but rather as a variation on conventional models of
primordial fluctuation generation (cfr. \cite{Peebles,BMPR,Starobinsky})
whereby quantum fluctuations of the inflaton field leave their imprint on
the primordial density field and are subsequently washed away. In other
words, the underlying physics in our model is the same as in these more
familiar approaches to primordial fluctuation generation: we do not question
the ultimate quantum origin of the fluctuations (another difference with the
cosmic turbulence theory from the fifties), but simply borrow insights from
hydrodynamics to describe the macroscopic behavior of these fluctuations,
rather than rely on possibly oversimplified linearized microscopic models.
The conditions of validity of our procedure are the assumptions that the
energy momentum tensor of fluctuations is a c-number quantity (which ought
to be true at any scale below Planck's) and the Second Law (which, unlike
the Third, has not been challenged yet to our knowledge).

A second goal of this paper is to demonstrate the application of the
hydrodynamic approach by discussing some simple solutions to the non linear
Navier - Stokes equation, and the resulting spectra of primordial
fluctuations. Since the general solution to the Navier- Stokes equation is
certainly not available, this requires the appeal to physical insights to
simplify the problem.

The basic mechanism of non linear hydrodynamic evolution is the energy
transport between eddies of different size through mode - mode coupling, and
the viscous dissipation of small scale eddies. The fundamental issue
regarding model building is to obtain a closed form energy balance equation,
which allows us to follow in time the shape of the energy spectrum. This
usually involves some closure hypothesis to reduce the infinite hierarchy of
dynamical equations for velocity correlation functions to a manageable set 
\cite{Schilling}. Lacking anything better, we shall fall back on the
time-honored hypothesis that the effect of smaller eddies at any given scale
may be simulated by a scale dependent effective viscosity \cite
{Batchelor,UFrisch,McComb}; for concreteness, we shall follow Heisenberg's
1948 formulation of this idea \cite{Hei48}.

The Heisenberg theory admits some very simple solutions with the property of
self-similarity. These have been worked out by Chandrasekhar \cite{Chandra}
and generalized to Friedmann - Robertson - Walker (FRW) backgrounds by
Tomita {\it et al }\cite{Tomita}. These solutions agree with the Kolmogorov
1941 theory in the inertial range \cite{McComb}, failing to reproduce
observations for very small eddies. Fortunately, we shall only require the
solution in the opposite limit of very large eddies, where it is trustworthy%
\footnote{%
We wish to point out that the applicability of Kolmogorov's spectrum to
large scale turbulence should not be taken for granted \cite{GCPRL}}. With
minor adjustments, Tomita's analysis of turbulence decay in FRW space times
also provides a solution to the evolution of our equivalent fluid.
Self-similarity is an appealing feature to us, as we do expect that a
generic flow will be eventually brought to some sort of steady state by the
inflationary expansion.

The task at hand is then to study the evolution of a typical eddy as it is
blown up by the universal expansion, exchanging and dissipating energy while
inside the horizon, and freezing when outside, until it reaches the
reheating hypersurface and delivers its energy to radiation. By assuming
that the turbulent velocity fluctuations in the eddy produce fluctuations in
the energy density of radiation in the usual way, we shall be able to relate
the primordial density contrast to the features of the original self similar
turbulence. The resulting spectrum may be matched against the known data on
the cosmic microwave background \cite{COBE}, providing a crucial test of the
inner consistency and viability of the approach. Our conclusion shall be
that, insofar as the horizon remains constant during inflation, the spectrum
of primordial density fluctuations produced by self similar flows is
strictly scale invariant ($n=1$, see \cite{HZ}) at large scales.
Quantitative agreement with observations may be obtained without any special
fine tuning.

A few comments on the possibility of deriving present day spectra from early
Universe hydrodynamics are in order; after all, one of the main criticisms
against the conventional cosmic turbulence theory was that, even if
turbulence were efficiently generated in the Early Universe, it would decay
and become uninteresting well before recombination. This criticism does not
apply to the present work, since we shall show that the results of customary
inflationary models and self similar flows concerning the primordial density
field immediately after reheating are identical. Further evolution of the
primordial density contrast after reheating is well described by the theory
of linear density fluctuations in an expanding Universe, a subject properly
covered by many textbooks (cfr. \cite{Peebles,BMPR,Weinberg,Börner}); the
result, for both the conventional inflationary models and the present ones,
is that these fluctuations in the primordial density contrast survive to
recombination time.

The key point in our analysis is the evolution of the flow during inflation,
as the different modes leave the horizon and ''freeze'', and we discuss this
issue at some length. The amount of turbulence at reheating is ample enough
to seed fluctuations at the level suggested by COBE data, provided the
initial temperature is large enough. The lower bound in the temperature,
however, is not so large that would invalidate the vacuum dominance
condition which is a presupposition of Inflation. Of course, whether
Inflation is likely to happen or not is a difficult question (cfr. \cite{me}%
) lying beyond the scope of this paper.

To conclude, the main objective of this paper is to show that it is possible
to develop sensible models of inflation where inflaton fluctuations evolve
nonlinearly and are very far from their vacuum states. The connection of the
physics of primordial fluctuations to hydrodynamics opens up a wealth of new
interesting phenomena, such as intermittency in the primordial spectrum \cite
{UFrisch,Frisch} and Burgers turbulence \cite{Burgers}, with a strong
potential impact on our understanding of the evolution of cosmic structures.
Moreover, it is appealing to be able to account for a macroscopic
phenomenon, such as fluctuation generation on super horizon scales, mostly
on macroscopic terms (for an independent attempt in this direction, see \cite
{zimdahl}). Most importantly, by not assuming that higher derivatives of the
potential are a priori negligible, we avoid a possible conflict within the
potential reconstruction program \cite{reconstruction}.

The rest of the paper is organized as follows. In next section we provide a
brief summary of hydrodynamics in flat and expanding universes, in order to
set up the language for the rest of the paper. In section III we proceed to
discuss the equivalent fluid description of inflaton fluctuations, and how
to extract the primordial density contrast therefrom. As a simple
application of the method, we consider briefly the case of free
fluctuations, showing that the model leads back to the conventional results.
In section IV we present Chandrasekhar's self-similar solutions and their
generalization to expanding universes, deriving the corresponding scale
invariant primordial contrast; we also discuss how relevant these solutions
are as actual descriptions of the inflationary period. We state our main
conclusions in the final section.

\section{Hydrodynamic flows}

\subsection{Flows in flat space time}

The equations governing the dynamics of a fluid in local thermodynamic
equilibrium are the continuity and Navier-Stokes ones, which, in the case of
flat space time, read:

\begin{equation}
\frac{\partial \rho }{\partial t}+\left( {\bf U\cdot }\nabla \right) \rho =0
\label{contpl}
\end{equation}

\begin{equation}
\frac{\partial {\bf U}}{\partial t}+\left( {\bf U\cdot }\nabla \right) {\bf %
U=-}\frac 1\rho \nabla p+\nu \nabla ^2{\bf U}  \label{NSpl}
\end{equation}
where we have assumed incompressibility, valid when typical velocities are
much smaller than the sound velocity; $\nu =\eta /\rho $ is the kinematic
shear viscosity. The transition from laminar to turbulent motion can be
universally described by the dimensionless ''Reynolds'' number:

\[
R=\frac{UL}\nu 
\]
where $U$ is a typical velocity and $L$ a typical length scale. This number
represents the order of magnitude of the ratio of the inertial to the
viscous term. Low Reynolds numbers correspond to laminar motion, while high
ones suggest turbulent behavior.

In general, the velocity profile displays variations in space and time. This
implies that the flow must be described probabilistically. Thus, each
quantity involved in (\ref{contpl}-\ref{NSpl}) is divided in its mean value
and a fluctuation from it; for example, we write ${\bf U}=\bar U+u$, where $%
u $ stands for the fluctuating part of the velocity. In the case where
motion is isotropic, the mean value $\bar U$ for the velocity must be zero,
since otherwise there would be a preferred direction.

To analyze the system's behavior, we define the two-point one-time
correlation function for the velocity:

\begin{equation}
R_{ij}({\bf x,x}^{\prime },t)=\left\langle u_i({\bf x,}t)u_j({\bf x}^{\prime
}{\bf ,}t)\right\rangle  \label{Rij}
\end{equation}
In the case of homogeneous and isotropic motion, this correlation must be
only a function of the time $t$ and the distance between ${\bf x}$ and ${\bf %
x}^{\prime }$, i.e. $R_{ij}({\bf x,x}^{\prime },t)=R_{ij}(r,t)$, where $%
r=\left| {\bf x-x}^{\prime }\right| .$ Observe that $R_{ii}(0,t)$ (summation
over repeated indices must be understood) is twice the average energy
density of the flow at time $t$. From (\ref{NSpl}) we obtain the equation
that this correlation must obey, namely:

\begin{equation}
\frac \partial {\partial t}R_{ij}(r,t)=T_{ij}(r,t)+P_{ij}(r,t)+2\nu \nabla
^2R_{ij}(r,t)  \label{ecRij}
\end{equation}
where 
\begin{equation}
P_{ij}(r,t)=\frac 1\rho \left( \frac \partial {\partial r_i}\left\langle p(%
{\bf x,}t{\bf )}u_j({\bf x}^{\prime },t)\right\rangle -\frac \partial {%
\partial r_j}\left\langle p({\bf x}^{\prime },t)u_i({\bf x},t)\right\rangle
\right)  \label{Pij}
\end{equation}
and

\begin{equation}
T_{ij}(r,t)=\frac \partial {\partial r_k}\left\langle u_i({\bf x},t)u_k({\bf %
x},t)u_j({\bf x}^{\prime },t)-u_i({\bf x},t)u_k({\bf x}^{\prime },t)u_j({\bf %
x}^{\prime },t)\right\rangle  \label{Tij}
\end{equation}
The tensor $T_{ij}$ comes form the inertia term in Navier-Stokes equation
and, as it involves a product of third order in the velocity, reflects the
fact that there is not a close set of equation for the correlations of
successive orders but there is a hierarchy of equations instead. The problem
of closing that hierarchy is known as the ''moment closure problem'' \cite
{Schilling}. Let us call $\Phi _{ij}(k,t)$ the Fourier transform of $%
R_{ij}(r,t)$. Then the energy density becomes

\[
\frac 12R_{ii}(0,t)=\int E(k,t)\;dk, 
\]
where

\begin{equation}
E(k,t)=\frac 12\int \Phi _{ii}({\bf k},t)\ k^2\ d\Omega ({\bf k)}  \label{ET}
\end{equation}

is the energy density stored in eddies of size $k^{-1}.$ Defining $\Gamma
_{ij}$ as the Fourier transform of $T_{ij}$, we obtain from (\ref{ecRij})
the equation of balance of the energy spectrum: 
\begin{equation}
-\frac \partial {\partial t}E(k,t)=T(k,t)+2\nu k^2E(k,t)  \label{ecesp}
\end{equation}
where 
\begin{equation}
\text{\qquad }T(k,t)=-\frac 12\int \Gamma _{ii}({\bf k},t)\ k^2\ d\Omega (%
{\bf k)}  \label{ET1}
\end{equation}

The inertia term $T(k,t)$ is the one that contains the mode-mode
interaction, and its effect is to drain energy form the more energetic modes
-typically the bigger ones- to the ones where there is major viscous
dissipation -the smaller ones-.

\subsection{Flows in expanding universes}

For a curved space-time, in particular that described by a Friedmann -
Robertson - Walker (FRW) background metric with zero spatial curvature ($%
ds^2=-dt^2+a^2(t)\left( dx^2+dy^2+dz^2\right) $), the generalization of the
above arguments has been considered by many authors \cite
{vonWeizsacker,Nariai,SMT,OzernoiyChibisov,NariaiyTanabe}. We follow Tomita 
{\it et al.}'s analysis \cite{Tomita}, in which they obtain the solution for
the energy spectrum in the case of homogeneous, isotropic and incompressible
turbulence.

In a generic space time, we describe fluid flow from the energy density $%
\rho $, pressure $p$ and four velocity $U$. The symmetries of the FRW
solution suggest using instead the commoving three velocity $u^i=U^i/U^0;$
if $U^i\ll U^0$ the flow is non relativistic, and if $\nabla {\bf u=}0,$ it
is incompressible (${\bf u}=(u^1,u^2,u^3)$). Later on, we shall also use the
physical three velocity $v=a(t)u.$

The corresponding continuity and Navier-Stokes equations for a
Robertson-Walker background are obtained by the condition of conservation of
the energy-momentum tensor \cite{Weinberg}. For a non relativistic
incompressible fluid, with shear viscosity $\eta =\nu \left( p+\rho \right) $
(but no bulk viscosity), these reduce to:

\begin{equation}
\frac{\partial \rho }{\partial t}+3\frac{\dot a}a\left( p+\rho \right) =0
\label{contcur}
\end{equation}
\begin{equation}
\frac{\partial {\bf u}}{\partial t}+\left[ \left( {\bf u\cdot }\nabla
\right) +\frac{\partial \ln \left( (p+\rho )a^5\right) }{\partial t}\right] 
{\bf u=-}\frac{\nabla p}{a^2\left( p+\rho \right) }+\frac 1{a^2}\nu \nabla ^2%
{\bf u}  \label{NScur}
\end{equation}
where we have assumed that $p+\rho $ depends only on time. For the physical
three velocity ${\bf v}$, the corresponding Navier-Stokes equation reads: 
\begin{equation}
\frac{\partial {\bf v}}{\partial t}+\left[ \frac 1a\left( {\bf v\cdot }%
\nabla \right) +\frac{\partial \ln \left( (p+\rho )a^4\right) }{\partial t}%
\right] {\bf v=-}\frac{\nabla p}{a\left( p+\rho \right) }+\frac 1{a^2}\nu
\nabla ^2{\bf v}  \label{NScurv}
\end{equation}
In obtaining (\ref{contcur})-(\ref{NScurv}) we have neglected possible
perturbations to the FRW metric. The corresponding equations considering
fluctuations in the metric ($g_{\mu \nu }=g_{\mu \nu }^0+h_{\mu \nu }$) have
been obtained by Weinberg \cite{Weinberg}. The continuity equation is not
corrected by gravitational perturbations, while in the Navier-Stokes
equation the metric fluctuations appear explicitly only within the shear
viscosity term. It can be demonstrated that these terms involving metric
fluctuations are negligible for scales that are inside the horizon \cite{GC}%
. For scales bigger than the Hubble radius, since dissipation through
viscosity is not effective anyway, we may still use the unperturbed Navier -
Stokes equation.

The operation of Fourier transforming in the case of a Robertson-Walker
cosmology is done in terms of commoving wave-numbers. In doing so, the
following equation for the energy spectrum is obtained: 
\begin{equation}
-\frac \partial {\partial t}E(k,t)=T(k,t)+2\left\{ \frac{\nu k^2}{a^2}+\frac{%
\partial \ln \left( (p+\rho )a^4\right) }{\partial t}\right\} E(k,t)
\label{ecespcur}
\end{equation}
where the relationship between $E(k,t)$ and $\Phi _{ij}(k,t)$ as well as
between $T(k,t)$ and $\Gamma _{ij}(k,t)$ is the same as that for a flat
space time, if we define $R_{ij}$ and $T_{ij}$ from correlations of physical
quantities, as follows: 
\begin{eqnarray}
R_{ij}(r,t) &=&a^2\left\langle u_i({\bf x,}t)u_j({\bf x+r},t)\right\rangle
,\   \label{ETcur} \\
\ T_{ij}(r,t) &=&a^2\frac \partial {\partial r_k}\left( \left\langle u_i(%
{\bf x},t)u_k({\bf x},t)u_j({\bf x+r},t)\right\rangle -\left\langle u_i({\bf %
x},t)u_k({\bf x+r},t)u_j({\bf x+r},t)\right\rangle \right)
\end{eqnarray}

\section{Equivalent fluid for inflaton fluctuations}

After establishing the basic necessary notions for the description of
hydrodynamic flows, our goal is to associate an equivalent fluid description
to inflaton fluctuations, and to derive the spectrum of primordial density
fluctuations at reheating therefrom. We shall discuss in the following
sections some non trivial instances of this method.

\subsection{The inflaton as a fluid}

To describe the inflaton field from the point of view of an equivalent
fluid, we need to obtain the energy density, pressure and velocity of this
fluid as functionals of the state of the field. To this end, our starting
point will be that in the rest frame of the fluid (quantities in this frame
being labelled by a curl), the field ought to be spatially constant

\begin{equation}
\widetilde{\nabla }\phi =0  \label{gradcero}
\end{equation}

To obtain the fluid four-velocity, we make a boost to the commoving frame.
Then, the boost's characteristic velocity will be the one we are seeking
for. By the condition (\ref{gradcero}) we obtain: 
\begin{equation}
u_i=-\frac{\partial _i\phi }{\dot \phi }  \label{ui}
\end{equation}
which is generalized to the covariant form 
\begin{equation}
u_\mu =-\frac{\partial _\mu \phi }{\sqrt{-\partial _\rho \phi \ \partial
^\rho \phi }}  \label{cuadriv}
\end{equation}
The energy density in the rest frame must be:

\[
\widetilde{\rho }=\frac 12\left( \frac{\partial \phi }{\partial t}\right)
^2+V(\phi ) 
\]
Using the Lorentz transformations with the four-velocity (\ref{cuadriv}) we
obtain the general form for the energy density: 
\begin{equation}
\rho =-\frac 12\partial _\rho \phi \ \partial ^\rho \phi +V(\phi )
\label{rho}
\end{equation}
Finally, we obtain the pressure imposing an equality between the
energy-momentum tensor for a perfect fluid (see for example \cite{Weinberg})
and that for a minimally coupled scalar field (see \cite{BirDv}). The
resulting pressure is: 
\begin{equation}
p=-\frac 12\partial _\rho \phi \ \partial ^\rho \phi -V(\phi )
\label{presion}
\end{equation}
Since the possibility of deriving a Navier - Stokes equation for the
equivalent fluid rests on the conservation of $T_{\mu \nu }$, in principle
only the whole inflaton field can be thus represented. However, under the
approximation that the homogeneous part of the inflaton essentially
contributes an effective cosmological constant, the background energy
momentum tensor $T_{0\mu \nu }=\Lambda g_{\mu \nu }$ is independently
conserved (even if $\Lambda $ were not constant, conservation fails only on
scales too large to be cosmologically relevant), and we can associate an
equivalent fluid to the inhomogeneous quantum fluctuations $\delta \phi $
alone. For this fluid, we find the physical velocity ($v^i=au^i=a^{-1}u_i$) 
\begin{equation}
v_k^i=k_{phys}^i\left( \frac{\delta \phi _k}{\dot \phi _0}\right)
\label{basoon}
\end{equation}
where $\phi _0$ is the homogeneous background. We interpret this equation to
mean that stochastic averages of the fluid velocity are to be identified
with (symmetric) quantum expectation values of the operator in the right
hand side \cite{BM,CH94}. As far as the equation of state is concerned, the
free energy for a massive scalar field in the high temperature limit ($T\gg
m $) \cite{Ramond} gives us the relationship between the pressure and the
energy density, which turns out to be that for radiation, $p=\frac 13\rho $.
This means that the energy density for this fluid redshifts proportional to $%
a^{-4}.$ As this result has been obtained for a flat space-time, it is valid
for scales smaller than the curvature radius. When scales are bigger than
the Hubble radius, which in turn takes place when the high temperature limit
is no longer valid, the field's equation of motion $\Box \phi +V^{\prime
}\left( \phi \right) =0$ turns out to be 
\[
\frac{d^2\phi _0}{dt^2}+3H\frac{d\phi _0}{dt}+V^{\prime }(\phi _0)=0 
\]
for the background field $\phi _0$ (where we have neglected spatial
derivatives) and 
\begin{equation}
\frac{d^2\left( \delta \phi \right) }{dt^2}+3H\frac{d\left( \delta \phi
\right) }{dt}+V^{\prime \prime }(\phi _0)\delta \phi =0  \label{ecfluct}
\end{equation}
for the fluctuation $\delta \phi $ (where we have neglected spatial
derivatives as well as non linear terms). The time derivative of the
equation for the background field gives us an equation of motion for $\dot 
\phi _0:$ 
\begin{equation}
\frac{d^2\dot \phi _0}{dt^2}+3H\frac{d\dot \phi _0}{dt}+V^{\prime \prime
}(\phi _0)\dot \phi _0=0  \label{ecfipto}
\end{equation}
where we have used the constancy of $H$ during inflation. Comparing (\ref
{ecfluct}) and (\ref{ecfipto}) we see that, as $\dot \phi _0$ and $\delta
\phi $ obey the same equation of motion, they must be related by: $\delta
\phi =\dot \phi _0\ f\left( {\bf r}\right) $, which implies that the ratio $%
\frac{\delta \phi }{\dot \phi _0}$ must be independent of time (cfr. \cite
{GP}). For our fluid description, this means that when scales are much
bigger than the Hubble radius, the velocity $u_i$ (equation (\ref{ui})) must
remain constant, which in turn means that the physical three velocity $%
v^i=au^i$ must redshift proportional to $a^{-1}$. As these scales are frozen
out because they are outside the horizon, they cannot interact among them or
be dissipated by viscosity. Thus, the Navier - Stokes equation (\ref{NScurv}%
) reduces to 
\begin{equation}
\frac{\partial {\bf v}}{\partial t}+\frac{\partial \ln \left( (p+\rho
)a^4\right) }{\partial t}\ {\bf v=}0  \label{NSvouthor}
\end{equation}
We obtain $v\propto a^{-1}$ when $\left( p+\rho \right) \propto a^{-3},$
corresponding to the equation of state of matter: $p=0$. Thus, when the
scales are well outside the horizon, our fluid behaves as pressureless dust,
which agrees with the well known prediction based on Virial's theorem for
the equation of state of a field undergoing oscillations \cite{p=0}, which
occurs at the final period of inflation. We must point out that the
hypothesis of incompressibility is no longer valid for an equation of state
of this type. Nevertheless, for scales bigger than the Hubble radius, which
cannot decay through non linear interaction or dissipation by viscosity,
equation (\ref{NSvouthor}) is still valid, regardless of the ratio of
typical velocities to the speed of sound.

\subsection{Transport coefficients}

The framework to obtain transport coefficients for our fluid is linear
response theory. In the limit of slowly variations in space and time of the
magnitudes involved in the equation of conservation for the energy-momentum
tensor, the system's response while it is slightly displaced from
equilibrium can be alternatively described by Navier-Stokes and continuity
equations as well as by equilibrium expectation values of correlation
functions. Matching these two descriptions, one obtains the Kubo formula for
the shear viscosity \cite{KadMart}:

\begin{equation}
\eta =\frac 16\lim_{w,k\rightarrow 0}\left[ \frac 1w\int dt\int d^3{\bf r}\
e^{i\left( {\bf k\cdot r}-wt\right) }\left\langle \left[ \pi _{ij}({\bf r}%
,t),\pi ^{ij}({\bf 0,}0)\right] \right\rangle _{eq}\right]  \label{eta}
\end{equation}
where $\pi _{ij}$ are the traceless spatial-spatial components of the
energy-momentum tensor: 
\[
\pi _{ij}=T_{ij}-\frac 13\delta _{ij}T_{\ k}^k 
\]
For a minimally coupled scalar field the commutator involved in (\ref{eta})
turns out to be: 
\begin{equation}
\left[ \pi _{ij}(x),\pi ^{ij}(0)\right] =\left[ \partial _i\phi (x)\partial
_j\phi (x),\partial ^i\phi (0)\partial ^j\phi (0)\right] -\frac 13\left[
\partial _i\phi (x)\partial ^i\phi (x),\partial _j\phi (0)\partial ^j\phi
(0)\right]  \label{conmut}
\end{equation}
These commutators can be evaluated from retarded and advanced Green
functions for a massive scalar field coupled to other fields as well as to
itself, which must satisfy: 
\[
\left( \Box +\Gamma \frac \partial {\partial t}+m^2\right)
G_{R,A}(x,x^{\prime })=\delta ^4(x-x^{\prime }) 
\]
where $\Gamma $ is the thermal width which comes from the field's
interactions. From these Green functions we obtain the Pauli-Jordan two
point function $G(x,x^{\prime })=\left\langle \left[ \phi (x),\phi
(x^{\prime })\right] \right\rangle $:

\[
G(x,x^{\prime })=-\frac{2i\Gamma }{\left( 2\pi \right) ^4}\int d^4p\frac{%
e^{-i\left( p_0(t-t^{\prime })-{\bf p\cdot (x-x}^{\prime })\right) }p_0}{%
\left[ (p_o-i\Gamma )^2-{\bf p}^2-m^2\right] \left[ (p_o+i\Gamma )^2-{\bf p}%
^2-m^2\right] } 
\]
The mean value for the anticommutator of the field $G_1(x,x^{\prime
})=\left\langle \left\{ \phi (x),\phi (x^{\prime })\right\} \right\rangle $
is obtained through the Kubo-Martin-Schwinger's theorem \cite{KMS}:

\[
G_1(x,x^{\prime })=-\frac{2i\Gamma }{\left( 2\pi \right) ^4}\int d^4p\frac{%
e^{-i\left( p_0(t-t^{\prime })-{\bf p\cdot (x-x}^{\prime })\right) }p_0\coth
\left( \beta p_o/2\right) }{\left[ (p_o-i\Gamma )^2-{\bf p}^2-m^2\right]
\left[ (p_o+i\Gamma )^2-{\bf p}^2-m^2\right] } 
\]
>From these, using Wick's theorem \cite{Wick} and the c-number character of
the mean value for the commutator, we obtain the commutator of the product
of two fields as: 
\[
\left\langle \left[ \phi (x)\phi (x),\phi (x^{\prime })\phi (x^{\prime
})\right] \right\rangle =\left\langle \left[ \phi (x),\phi (x^{\prime
})\right] \right\rangle \left\langle \left\{ \phi (x),\phi (x^{\prime
})\right\} \right\rangle 
\]
The first term involved in (\ref{conmut}) turns then out to be 
\[
\int dt\ d^3{\bf r}\ \ e^{i\left( {\bf k\cdot x-}wt\right) }\left\langle
\left[ \partial _i\phi (x)\partial _j\phi (x),\partial ^i\phi (0)\partial
^j\phi (0)\right] \right\rangle =\ 
\]
\[
\frac{-4\Gamma ^2}{\left( 2\pi \right) ^8}\int d^4p\frac{\left\{ {\bf p}^2(%
{\bf k}-{\bf p})^2+\left[ {\bf p\cdot }({\bf k}-{\bf p})\right] ^2\right\}
\left( w-p_o\right) p_0\coth \left( \beta p_o/2\right) }{\left| (p_o-i\Gamma
)^2-{\bf p}^2-m^2\right| ^2\left| (w-p_o-i\Gamma )^2-{\bf p}^2-m^2\right| ^2}%
\label{vi} 
\]
$\label{vi}$and a similar formula for the second term in (\ref{conmut}).
This gives for the shear viscosity in the high temperature limit ($T\gg
m,\Gamma $): 
\begin{equation}
\eta =const\ T\ \Gamma ^2  \label{etagama}
\end{equation}
The thermal width $\Gamma $ comes from the imaginary part of the
self-energy. From dimensional analysis, it must be proportional to the
temperature since the only relevant scale in the high temperature limit is
the temperature itself (the constant of proportionality must be much less
than unity for the consistency of the high temperature limit). Assuming that
the only present interaction is the one coming from a $\sigma \phi ^4$ term, 
$\Gamma $ must include two $\sigma $ insertions, which means that $\Gamma $
must be proportional to $\sigma ^2$ (we will estimate it as $\Gamma \sim
\sigma ^2T$). The shear dynamic viscosity becomes 
\begin{equation}
\eta \sim \sigma ^4T^3  \label{etaT}
\end{equation}
A similar analysis allows us to evaluate the bulk viscosity, which turns out
to be zero for a fluid with an equation of state of the type $p=\frac 13\rho 
$ \cite{YS}, in agreement with our previous assumptions.

\subsection{Conversion of hydrodynamic fluctuations into primordial density
contrast}

Having described the quantum field as a fluid, we will analyze the resulting
spectrum of density inhomogeneities. To do so, we assume that at some time $%
t_1$ during the beginning of the inflationary phase, when all the scales
relevant to cosmology were inside the horizon, the scalar field fluctuations
were undergoing hydrodynamic fluctuations. Once the scales leave the Hubble
radius, their energy cannot be dissipated by viscosity or by nonlinear
coupling. Thus, equation (\ref{ecespcur}) means that they evolve according
to:

\begin{equation}
E(k,t>t_{out})=E(k,t=t_{out})\left[ \frac{\left( (p+\rho )a^4\right) _{out}}{%
\left( (p+\rho )a^4\right) (t)}\right] ^2  \label{Eaf}
\end{equation}
where the subscript $"out"$ refers to the time when each scale leaves the
horizon.

The definition of $E(k)$ (equation (\ref{ET})) can be written in terms of
the Fourier transform of the velocity; since the flow is statistically
isotropic and homogeneous:

\begin{equation}
\left\langle v^i({\bf k)}v^i({\bf k}^{\prime })\right\rangle =\left( \frac{%
E(k)}{4\pi k^2}\right) \delta ^3({\bf k+k}^{\prime })  \label{viola}
\end{equation}

Combining (\ref{Eaf}) and (\ref{viola}) we can obtain the r.m.s. value for
the scalar field velocity at the time of reheating, which will be the r.m.s.
value of the perturbation in the radiation's streaming velocity. This
perturbation will in turn produce fluctuations in the energy density of
radiation, which will evolve in the usual way. The theory of relativistic
very large wavelength fluctuations predicts $\delta \sim a^2$, where $\delta
=\delta \rho /\rho $ is the density contrast, and thus $\dot \delta \sim
H\delta $, while the continuity equation yields $\dot \delta \sim v/l$ on a
scale of physical size $l$. Consistency of these two pictures leads to the
relationship between the velocity at the time of reheating and the
fluctuation in the energy density as

\begin{equation}
\left. \frac{\delta \rho }\rho \right| _{reh}=\frac{v_{reh}}{lH_{reh}}
\label{B}
\end{equation}
where $H_{reh}$ is the Hubble parameter at reheating. Following this
fluctuations up to the time they reenter the Hubble radius, assuming that
their size is such that they are always unstable (they must be always bigger
than the Jeans's length), they grow following the law (see for example \cite
{Peebles,Weinberg}):

\begin{equation}
\delta \equiv \left. \frac{\delta \rho }\rho \right| _{ent}=\left. \frac{%
\delta \rho }\rho \right| _{reh}\left( \frac{a_{eq}}{a_{reh}}\right)
^2\left( \frac{a_{ent}}{a_{eq}}\right) \equiv H_{reh}^2\ l^2\left. \frac{%
\delta \rho }\rho \right| _{reh}  \label{C}
\end{equation}
where the subscript $"ent"$ means the time each scale reenters the Hubble
radius (the second equation holds even if the entering time occurs before
matter-radiation equality). Combining equations ((\ref{B})-(\ref{C})), we
obtain the density contrast predicted by this theory at the time the modes
reenter the Hubble radius:

\begin{equation}
\left\langle \delta _k\delta _{k^{\prime }}\right\rangle _{ent}=\frac{%
H_{reh}^2a_{reh}^2E(k,t=t_{reh})}{4\pi k^4}\ \delta ^3({\bf k+k}^{\prime })
\label{derhoderho}
\end{equation}
This is the main result of this paper, as it relates the density contrast to
a hydrodynamic variable. We shall see a nontrivial application of this
formula in next section, but before, it is convenient that we pause to show
explicitly how the familiar results relating to free field fluctuations are
recovered in this language. Probably the most important feature of a theory
where inflaton fluctuations are free is that each mode evolves independently
of the other ones. Immediately after leaving the horizon they freeze, a
situation that can be described phenomenologically assigning to the mode the
effective equation of state of dust. This implies that 
\begin{equation}
E(k,t=t_{reh})=E(k,t=t_{out})\left( \frac{a(t=t_{out})}{a(t_{reh})}\right) ^2
\label{cuatro}
\end{equation}
\begin{equation}
\left\langle \delta _k\delta _{k^{\prime }}\right\rangle _{ent}=\frac{%
E(k,t=t_{out}(k))}{4\pi k^2}\ \delta ^3({\bf k+k}^{\prime })  \label{cinco}
\end{equation}
Let us compare this expression to the usual one in terms of quantum
fluctuations. First we use Eq. (\ref{viola}), neglecting any variation of $H$
or of the velocities during reheating, to get

\begin{equation}
\left\langle \delta _k\delta _{k^{\prime }}\right\rangle _{ent}=\ \left.
\left\langle \ v^i({\bf k)}v^i({\bf k}^{\prime })\right\rangle \right|
_{t=t_{out}(k)}  \label{cello}
\end{equation}
We now relate the physical velocity to field fluctuations according to Eq. (%
\ref{basoon}); at $t=t_{out}(k)$, $k_{phys}=H$, and this reduces to 
\begin{equation}
\left\langle \delta _k\delta _{k^{\prime }}\right\rangle _{ent}=\left( \frac 
H{\dot \phi }\right) ^2\left. \left\langle \delta \phi _k\delta \phi
_{k^{\prime }}\ \right\rangle \right| _{t=t_{out}(k)}  \label{oboe}
\end{equation}
which is the conventional result \cite{GP}.

This shows the agreement between the fluid description and the conventional
approach in this case, although of course it is only in the nonlinear case
where we expect the hydrodynamic formalism to bring definite advantages.

\section{Self similar flows and nonlinear fluctuations}

In the previous sections we set up the general formalism whereby we can
associate to the evolution of quantum fluctuations during inflation an
equivalent fluid description, and derive the corresponding primordial
density contrast from hydrodynamic variables. Of course, to put the
formalism to actual use, we must be able to solve the Navier - Stokes
equations, which is in itself almost as daunting as solving the fundamental
quantum field theory. However, there is in the hydrodynamic case a century
of lore to draw upon \cite{lamb}, and some well tested approximations
leading to relatively simple solutions. In this section, we shall
demonstrate the equivalent fluid method by investigating the spectra
resulting from one of these solutions, namely self similar flows. Towards
the end of the section, we shall discuss the relevance of these solutions to
actual cosmology.

\subsection{Self similar flows in flat and expanding universes}

As we have seen in the previous section (Eq.(\ref{derhoderho})), the key
element in deriving the primordial density contrast is the energy spectrum $%
E(k)$ (Eq.(\ref{ET})), which is the solution of the balance equation (Eq.(%
\ref{ecesp})). In it, the right hand side contains the viscous dissipation
as well as the inertial force $T(k,t)$. The overall effect of this term is
to transfer energy from a given scale to smaller ones through mode - mode
coupling; thus it is natural to model the action of the inertia term as a
source of viscous dissipation, where the effective turbulent viscosity for a
given mode depends on the motion of all smaller eddies \cite{UFrisch}. By
providing closure, that is, writing this effective viscosity in terms of the
spectrum itself, a closed evolution equation for $E(k)$ is obtained.
Concretely, Heisenberg \cite{Hei48} proposed the ansatz

\begin{equation}
\int_0^kT(k^{\prime },t)\ dk^{\prime }=2\nu (k,t)\int_0^kE(k^{\prime },t)\
k^{\prime \ 2}\ dk^{\prime }  \label{Heis1}
\end{equation}
where

\begin{equation}
\nu (k,t)=A\int_k^\infty \sqrt{\frac{E(k^{\prime },t)}{k^{\prime 3}}}\
dk^{\prime }  \label{Heis2}
\end{equation}
and $A$ is a dimensionless constant. With this hypothesis (known as the
Heisenberg hypothesis) as the solution to the closure problem, Chandrasekhar 
\cite{Chandra} has obtained the energy spectrum for decaying turbulence,
assuming that there is a stage in the decay where the bigger eddies have
sufficient amount of energy to maintain an equilibrium distribution, thus
requiring that the solution for the spectrum should be self-similar. With
this consideration into account he obtained an energy spectrum:

\begin{equation}
E(k,t)=\frac 1{A^2k_0^3t_0^2}\sqrt{\frac{t_0}t}F\left( \frac{k\sqrt{t}}{k_0%
\sqrt{t_0}}\right) \qquad
\end{equation}
where $k_0$ and $t_0$ are initial conditions (namely, the wave number
corresponding to the bigger eddy and its typical time of evolution). The
function $F$ obeys the equation

\begin{equation}
\frac 14\int_0^x\left[ F(x^{\prime })-x^{\prime }\frac{dF(x^{\prime })}{%
dx^{\prime }}\right] dx^{\prime }=\left\{ \nu k_0^2t_0+\int_x^\infty \frac{%
\sqrt{F(x^{\prime })}}{x^{\prime \ 3/2}}dx^{\prime }\right\}
\int_0^xF(x^{\prime })x^{\prime \ 2}dx^{\prime }  \label{ecF}
\end{equation}
which predicts a Kolmogorov type behavior for an inviscid fluid ($%
R\rightarrow \infty $, $R=\frac 1{\nu k_0^2t_0}$) in the ultraviolet limit:

\[
F(x)\rightarrow const\ x^{-5/3}\ (\nu =0\ ,\ x\rightarrow \infty )\label
{compkol} 
\]
While for nonzero viscosity:

\[
F(x)\rightarrow const\ x^{-7}\ (\nu \neq 0\ ,\ x\rightarrow \infty )\label
{comp-7} 
\]

In the infrared limit, $F$ has the universal behavior $F(x)=4x$ $(x\ll 1)$,
and thus we find a linear energy spectrum for $k\sqrt{t}\ll k_0\sqrt{t_0}.$
Chandrasekhar's self similar solutions are easily generalized to flows in
expanding Universes. The dependence on time and wave-number for the self
similar energy spectrum is \cite{Tomita}

\begin{equation}
E(k,t)=v_{ti}^2\left( \frac{(p+\rho )_ia_i^4}{(p+\rho )a^4}\right) ^2\frac{%
\lambda _i^2}\lambda F(\lambda k)  \label{speccur1}
\end{equation}
where the subscript $i$ refers to ''initial'' and $\lambda $ and $v_t$ are
respectively the Taylor's microscale and an average turbulent velocity,
defined as:

\begin{equation}
\lambda ^2(t)\equiv 5\frac{\int E(k,t)\ dk}{\int E(k,t)\ k^2\ dk}\ \qquad 
\frac 12v_t^2(t)\equiv \int E(k,t)\ dk  \label{lambdayvt}
\end{equation}
Their time evolution must follow the law: 
\begin{equation}
\lambda ^2(t)=\lambda _i^2+10\int_{t_i}^t\frac \eta {(p+\rho )a^2}\ dt\qquad
v_t=v_{ti}\left( \frac{(p+\rho )_ia_i^4}{(p+\rho )a^4}\right) \frac{\lambda
_i}{\lambda \left( t\right) }  \label{l(t)yv(t)}
\end{equation}
The viscosity for our fluid, at least at high temperature, is given by Eq. (%
\ref{etaT}). The equation which determines de function $F(\lambda k)$ in (%
\ref{speccur1}) turns out to be the same as in flat space time, Eq. (\ref
{ecF}), which means that assuming Heisenberg's hypothesis the spectrum is
linear in $k$ for scales much bigger than the Taylor's microscale \footnote{%
We wish to point out an ambiguity concerning the meaning of Heisenberg's
hypothesis in the case of curved spaces. For flat space time, the
proportionality between the integral up to a certain wave number $k$ of the
inertia and the viscous forces is given by (\ref{Heis1}) and (\ref{Heis2}).
In the case of a FRW space time, the autosimilar solution required by Tomita 
{\it et al.} (\ref{speccur1}) needs a time dependent dimensionless constant $%
A$ proportional to $\eta a^2$ for the consistency of the solution. This
product does remain constant only if the dynamic shear viscosity evolves in
time proportional to $a^{-2}.$ Thus unless this is the case, the solution we
described looks like a natural curved space generalization of the Heisenberg
- Chandrasekhar solution, but does not admit the same physical
interpretation.}: 
\begin{equation}
E(k,t)=4v_{ti}^2\left( \frac{(p+\rho )_ia_i^4}{(p+\rho )a^4}\right)
^2\lambda _i^2k\qquad \text{for \ }\lambda k\ll 1  \label{speccur}
\end{equation}

\subsection{Nonlinear inflationary models}

We now want to place a self similar solution in the context of a
inflationary scenario where, instead of regarding the inflaton fluctuations
as free, we shall substitute them by an equivalent fluid, whose evolution we
will assume to be self similar. We discuss at the end whether this last
assumption is a reasonable one.

As before, we will assume a duration of inflation close to the minimum value
($N_{\min }\simeq 60$, where $N$ stands for the number of {\it e-folds), }%
which can be justified by the expected quadrupole anisotropy \cite
{Grishchukdur} as well as by the ratio of the present to the critical
density (see Appendix). By this assumption, a scale whose present size
equals the horizon ($\simeq 3000$ $Mpc$) leaves the Hubble radius soon after
the beginning of inflation.

Unless in the free field case, here we cannot deal with each mode
independently, but we must treat the whole flow subject to a
phenomenological equation of state. Let us assume the self similar flow sets
in at a time $t_1$ when the temperature $T>>H$ (we discuss whether this is a
suitable assumption below); and that the present horizon scale leaves the
horizon at or around time $t_1$. Then it is valid to use the high
temperature limit for length scales close to the present horizon while they
leave the Hubble radius during the inflationary phase. The fluid's equation
of state in this limit is of the $p=\frac 13\rho $ type, which means that
the product $\left( p+\rho \right) a^4$ remains constant throughout the
universal expansion. The factor $\left( (p+\rho )a^4\right) _{out}$ involved
in (\ref{Eaf}) is then independent of the particular scale being considered
within this group. Thus, by (\ref{Eaf}) and (\ref{speccur}) we can obtain
the energy spectrum for these scales while they are outside the horizon:

\begin{equation}
\frac{E\left( k,t>t_{out}(k)\right) }{E\left( k_0,t>t_{out}(k_0)\right) }=%
\frac{E\left( k,t=t_{out}(k)\right) }{E\left( k_0,t=t_{out}(k_0)\right) }
\label{spec}
\end{equation}
where (cfr. (\ref{speccur}))

\begin{equation}
E(k,t=t_{out})=4v_t^2(t_1)\lambda ^2(t_1)k  \label{tuba}
\end{equation}
$\lambda (t_1)$ being the commoving Taylor's microscale at the time the self
similar flow sets in. As at the initial time $t_1$ the only relevant scale
is the temperature, we expect the initial Taylor's microscale to be the
inverse of the temperature at that time, i.e. $\lambda _{phys}(t_1)\sim 
\frac 1{T(t_1)}.$ By (\ref{l(t)yv(t)}) we can obtain the commoving Taylor's
microscale at later times, such that most of the scales are still inside the
horizon:

\begin{equation}
\lambda ^2(t)=\lambda _{\ }^2(t_1)+10\int_{t_1}^t\frac \eta {\left( p+\rho
\right) a^2}dt\simeq \lambda _{\ }^2(t_1)+10\frac{\eta (t_1)}{\left( p+\rho
\right) (t_1)}\frac 1{Ha^2(t_1)}  \label{lambda(t)}
\end{equation}
where we have used the fact that for a viscosity dependence upon time given
by (\ref{etaT}) the main contribution to the integral is given by its lower
limit, provided there is a difference between the times $t$ and $t_1$ of
more than an e-fold. Since 
\[
\frac{\eta (t_1)}{\lambda ^2(t_1)H\left( p+\rho \right) (t_1)}\sim \sigma ^4%
\frac{T(t_1)}H 
\]
this means that the commoving Taylor's microscale freezes in its initial
value, unless coupling is very strong. Let us take as reference scale $k_0$
in Eq. (\ref{spec}) the inverse Taylor's microscale (which is the last scale
to leave the horizon in the high temperature regime). Then, from Eq. (\ref
{Eaf})

\begin{equation}
\frac{E\left( k_0,t_{reh}\right) }{E\left( k_0,t=t_{out}(k_0)\right) }\sim
\left( \frac{a(T_{phys}=H)}{a(t_{reh})}\right) ^2  \label{horn}
\end{equation}

From Eqs. (\ref{spec}) and (\ref{tuba}), 
\begin{equation}
E\left( k,t_{reh}\right) \sim \left( \frac{a(T_{phys}=H)}{a(t_{reh})}\right)
^24v_t^2(t_1)\lambda ^2(t_1)k\sim \left( \frac{2v_t(t_1)}{a(t_{reh})H}%
\right) ^2k  \label{fiddle}
\end{equation}
and so from Eq. (\ref{derhoderho}) 
\begin{equation}
\left\langle \delta _k\delta _{k^{\prime }}\right\rangle _{ent}=\frac{%
v_t^2(t_1)}\pi \frac 1{k^3}\ \delta ^3({\bf k+k}^{\prime })
\label{derhoderho2}
\end{equation}
That is, a scale invariant Harrison - Zel 'dovich spectrum\cite{Berts} with
amplitude $v_t.$

On the other hand, let us recall that the fluid velocity is related to the
underlying field description by $v\sim k_{phys}\delta \phi /\dot \phi $. Now
the typical wave number is $k_{phys}\sim T$, and at high temperature also $%
\delta \phi \sim T.$ From the background evolution, we have $\dot \phi \sim
V^{\prime }(\phi )/3H\sim m_p^2H/3\phi .$ The number of e-foldings $N\sim
H\phi /\dot \phi \sim \phi ^2/m_p^2,$ so $\phi \sim \sqrt{N}m_p$, $\dot \phi
\sim m_pH$, and 
\begin{equation}
v_t(t_1)\sim \frac{T^2(t_1)}{m_pH}  \label{chiripa}
\end{equation}
Since the ratio is necessarily less than one, the hypothesis of a non
relativistic incompressible flow is seen to be consistent. The constraint of 
$k\ll \lambda ^{-1}$ reduces to a minimum scale above which we obtain scale
invariance. Jeans's length imposes a lower limit bigger than this (we are
assuming that the scales are always unstable while they are outside the
horizon, which is valid if they are bigger than the Jeans's length).

Finally, combining the estimates for the initial Taylor's microscale and the
turbulent velocity, we obtain a Reynolds number: 
\[
R=\frac 43\frac{\lambda _{phys}(t_1)v_t(t_1)\rho (t_1)}{\eta (t_1)}\sim 
\frac{v_t(t_1)}{\sigma ^4} 
\]
suggesting highly turbulent motion, specially for small couplings (the self
similar solutions are not dependent on high Reynolds numbers anyway). The
estimate of the viscosity must be taken with a grain of salt, nevertheless,
since we have ignored non perturbative effects \cite{YS}.

Observations suggest that density fluctuations have a scale invariant
spectrum with an amplitude $\delta \rho /\rho \sim 10^{-5}$ over a range of
scales going from maybe as low as $100Mpc$ up to the present horizon scale
at $l_{now}=3000Mpc\sim 10^{41}GeV^{-1}$\cite{COBE} \cite{Einasto}. Recall
Eq. (\ref{app1}) from the Appendix, relating the size of the Universe at the
time $t_{exit}$ when the actual horizon's scale left the inflationary
horizon, to the Hubble parameter during inflation. At the time $t_{exit}$,
the physical Taylor microscale is at most $\lambda =\left(
a(t_{exit})/a(t_1)\right) T^{-1}(t_1)=10^{-2}H^{-1}$, since a larger value
would narrow too much the regime where the spectrum is scale invariant. So

\[
\frac{T(t_1)}{\sqrt{m_pH}}=\left( \frac{a(t_{exit})}{a(t_1)}\right)
10^2\left( \frac H{10^{19}GeV}\right) ^{1/2}=10^{-26}\left( \frac{a(t_f)}{%
a(t_1)}\right) 
\]
We obtain agreement with observations (cfr.(\ref{derhoderho2}) and (\ref
{chiripa})) provided

\[
\left( \frac{a(t_f)}{a(t_1)}\right) \sim 10^{23}=e^{53} 
\]

Since the total run of inflation is some sixty e-foldings (no more than $70$
even in the extreme case $H=m_p$) this is not hard to obtain, provided self
similarity sets in a few e-foldings after the beginning of inflation.

We have thus set an explicit model where interacting inflaton fluctuations
lead to a density contrast in agreement with observations. We could rest our
case at this point, but before that, we would like to discuss briefly how
likely it is that this solution was actually realized in the Early Universe.

\subsection{Self similar flows and our Universe}

As we have seen above, a self similar turbulent flow pattern in the
equivalent fluid could explain the scale invariant spectrum of primordial
density fluctuations observed at scales above $100Mpc$, provided the self
similar regime sets early enough. Our goal in this final section is to
discuss whether this is a likely assumption regarding our own Universe.

In general, it is known that turbulent flows tend to relax towards self
similarity, but it is difficult to estimate how long does it take to get
there. In a rough approximation, there are two issues involved, first, on
what time scale the Heisenberg closure condition becomes valid, and then,
how long does it take the closed equation for the spectrum to yield self
similarity.

If posed in these terms, we may observe the analogy with the problem of
equilibration in the theory of dilute gases. In the latter, the fundamental
description of the dynamics is the BBGKY hierarchy\cite{Huang}. However,
after a time of the order of a collision time the hierarchy can be closed,
turning into the Boltzmann equation, which in turn leads to equilibrium in
times of the order of the mean free time.

In the turbulence problem, the time scale of a typical eddy is 
\begin{equation}
\tau =\frac \lambda {v_t}=\frac{m_pH}{T^3}\sim \frac 1H\left( \frac H{Tv_t}%
\right)  \label{amapola}
\end{equation}

We may take this as an estimate of the mean free time. The collision time
can be estimated as the characteristic time for the smallest eddies in the
flow (we visualize a collision as the exchange of a small eddy between two
larger ones). According to Landau - Lifshitz \cite{Landaulif}, this is 
\begin{equation}
\tau _{\min }=\tau R^{-3/4}\ll \tau  \label{azucena}
\end{equation}

It is simple to find parameters leading to $\tau $ of the order of an
e-folding, and therefore $\tau _{\min }\ll H^{-1}$; for example, take $%
H=10^6GeV$, $T(t_1)=10^{10}GeV,$ $v_t(t_1)=10^{-5}$, with $%
T_{reh}=10^{12}GeV $. These solutions may not be available in models with
such simple potentials as $\sigma \phi ^4$, but they are feasible in
multiparameter models such as those invoked in the potential reconstruction
program \cite{reconstruction}. While this does not constitute a proof, it
makes it plausible that a self similar solution may be realized in the early
stages of inflation.

Once the motion becomes self-similar, it may last for a rather long time.
From the equation of motion Eq. (\ref{ecespcur}) we expect the energy in the
flow to be dissipated in a time scale of the order of $\tau _{rel}=\nu
^{-1}\lambda _{phys}^2$. With $\lambda \sim T^{-1}$ and $\nu \sim \eta
/\left( \rho +p\right) \sim \sigma ^4T^{-1}$, we obtain

\begin{equation}
\tau _{rel}\sim \frac 1{\sigma ^4T}\sim R\tau  \label{trel}
\end{equation}

This time will be generally larger than an e-folding, and might be even
larger than the whole duration of inflation if $R$ is large enough.

We conclude that self similarity will be easily achieved for scales around
the Taylor microscale or smaller, and will propagate to larger scales after
several e-foldings. At larger scales there could be deviations from scale
invariance, associated with the transient behavior of the flow. We may
conjecture that as the fluid cools down typical velocities will decrease,
and thereby the primordial density contrast too. Thus the model will
naturally yield higher power at larger scales, reproducing the behavior of
multiple inflation models \cite{staropol}.

It bears mention that the same situation occurs in the usual treatments of
the free field case. From Eq. (\ref{oboe}), we may conclude that the
spectrum of primordial density fluctuations has an amplitude $\delta _k\sim
(H^2/\dot \phi )\sqrt{1+n_k}$, where $n_k$ is the occupation number for that
mode in the initial state of the inflaton field. Vacuum dominance requires $%
n_k\ll 1$ for $k\geq \sqrt{m_pH}$, but places no real restriction on larger
scales which, in minimal models of inflation, are still observable. Thus we
only obtain the conventional result (corresponding to $n_k=0$ for all modes)
for specially chosen initial conditions.

\section{Final remarks}

In present conventional approaches, density inhomogeneities arise from
primordial fluctuations in the inflaton field, ultimately of quantum origin.
Fluctuations are treated as a free field, thus forcing upon us the
assumption that higher derivatives of the inflaton potential are negligible.
In this paper we sought a direct estimate of the primordial density contrast
generated in a nonlinear inflationary model. Instead of assuming
fluctuations to behave as a free field, we consider them to be coupled, so
that they can be described phenomenologically as a fluid. We showed that
there are flow patterns for this fluid that reproduce observations at very
large scales.

While our model takes advantage of many insights from earlier studies of the
role of turbulence in structure formation (see \cite{GC,Tomita} etc.), it
explores a whole new aspect of the problem in the sense that it places
turbulence at the origin of the primordial fluctuations, rather than being
excited from them. It is therefore free from the criticisms that are usually
raised against the turbulence theory of structure formation \cite
{PeeblesLSS,Raychaudhuri}. To the contrary, our model successfully
reproduces the results of the conventional approach on very large scales,
that is, a scale invariant spectrum with a density contrast of about $%
10^{-5} $. While it is not completely free of fine tuning, the fine tuning
involved is the same as in the usual scenario.

In our view, the main result of our work is not that a self similar solution
should be our final description of fluctuations during inflation, but rather
that it is possible to make sense of the physics of fluctuations even in
rather general potentials. The self similar solutions we have explored in
some detail should be seen as an ideal case which will more or less
approximate actual flow patterns; indeed, the same could be said of the De
Sitter invariant vacuum as a description of the actual state of the field in
free theories.

The connection of hydrodynamics to fluctuation generation has some interest
of its own, as it provides an alternative to brute force quantum field
theoretic calculations, and also yields physical insight on the macroscopic
behavior of quantum fields in the Early universe. The equivalent fluid
method may be used to advantage also in other regimes, such as the pre -
inflationary Universe and the reheating era, where the strong back reaction
phase has so far been untractable \cite{BdV,Steve}. Moreover, it opens up a
wealth of new phenomena, such as intermittence \cite{Frisch,UFrisch} and
shocks \cite{Burgers}, which are not apparent in the customary treatments.
We will continue our research in this field, which promises a most rewarding
dialogue between cosmology, astrophysics, and nonlinear physics at large.

\section{Acknowledgments}

We are grateful to C. Ferro Font\'an and F. Minotti for pointing out refs. 
\cite{Schilling,GCPRL} to us, and to D. G\'omez for multiple conversations
on turbulence theory.

This work has been partially supported by the European project CI1-CT94-0004
and by Universidad de Buenos Aires, CONICET and Fundaci\'on Antorchas.

\section{Appendix}

The minimal amount of inflation necessary to solve the homogeneity problem
is obtained by the condition that a scale of the size of the horizon at
present ($\sim 3000Mpc$) should have been inside the Hubble radius at the
beginning of inflation. The Hubble radius during inflation is approximately
constant. A scale whose physical size at present is $\lambda (t_0)$ was, at
the end of reheating

\[
\lambda (t_{reh})=\lambda (t_0)\frac{a(t_{reh})}{a(t_0)}=\lambda (t_0)\frac{%
T(t_0)}{T_{reh}}\simeq \lambda (t_0)\ 2.35\cdot 10^{-25}\left( \frac{%
10^{12}GeV}{T_{reh}}\right) 
\]

Write $T_{reh}=\Gamma \sqrt{m_pH}$, where $\Gamma \leq 1$ and $H$ is the
Hubble parameter during inflation. Then towards the end of inflation we have

\[
\lambda \left( t_f\right) =\lambda (t_0)2.35\cdot 10^{-25}\left( \frac{%
10^5GeV}H\right) ^{1/2} 
\]

{\em \ }This scale left the horizon at a time $t_{exit}$ with $\lambda
\left( t_{exit}\right) =(a(t_{exit})/a(t_f))\lambda \left( t_f\right)
=H^{-1} $. So

\[
1=\left( \frac{a(t_{exit})}{a(t_f)}\right) \lambda (t_0)2.35\cdot
10^{-20}GeV\left( \frac H{10^5GeV}\right) ^{1/2} 
\]

In particular, for the present horizon scale we get

\begin{equation}
1=\left( \frac{a(t_{exit})}{a(t_f)}\right) 10^{21}\left( \frac H{10^5GeV}%
\right) ^{1/2}  \label{app1}
\end{equation}

Therefore, defining $N_{\min }=\ln \left[ a(t_f)/a(t_{exit})\right] $, which
would make $t_{exit}=t_i$ for the scale of the horizon at present

\begin{equation}
N_{\min }=\ln \left[ 10^{21}\left( \frac H{10^5GeV}\right) ^{1/2}\right]
=64.4+\ln \sqrt{\frac H{m_p}}  \label{app2}
\end{equation}
{\it \ }On the other hand, inflation could not have lasted much more than
this because otherwise the present density should be so fine tuned to the
critical one that would contradict observations as well as most speculations
on dark matter's density. This can be seen from one of the Friedmann's
equations:

\[
\Omega (t)-1=\frac k{\left( a(t)H(t)\right) ^2} 
\]
where $\Omega $ is the ratio of the density to the critical one and $k$ is
the spatial curvature of the FRW metric. This means that the ratio (assuming
instantaneous reheating):

\begin{equation}
\frac{\Omega (t_0)-1}{\Omega (t_i)-1}=\frac{\left( a(t_i)H(t_i)\right) ^2}{%
\left( a(t_0)H(t_0)\right) ^2}\simeq \exp 2\left[ 69.06+\ln \sqrt{\frac H{m_p%
}}-N\right]  \label{app3}
\end{equation}
So, if we set $\Omega (t_i)$ to lie in the interval $\left( 0,1\right) $, we
conclude that if $\Omega (t_0)$ is not fine tuned to $1$, $N$ should not
have been much more than $60.$

There is another argument supporting that inflation should not have spanned
much more than its minimum duration based on the comparison between the
expected quadrupole anisotropy and the detected one (see \cite{Grishchukdur}%
).

\end{document}